\documentclass[conference]{IEEEtran}
\IEEEoverridecommandlockouts
\usepackage{cite}
\usepackage{amsmath,amssymb,amsfonts}
\usepackage{algorithmic}
\usepackage{graphicx}
\usepackage{textcomp}
\usepackage{xcolor}
\usepackage{hyperref}
\usepackage{amsmath}

\def\BibTeX{{\rm B\kern-.05em{\sc i\kern-.025em b}\kern-.08em
    T\kern-.1667em\lower.7ex\hbox{E}\kern-.125emX}}
\begin{document}

\title{Automated Duplicate Bug Report Detection in Large Open Bug Repositories}

\author{
\IEEEauthorblockN{Clare E. Laney\textsuperscript{*}}
\IEEEauthorblockA{\textit{Department of Computer Science} \\
\textit{University of Colorado Colorado Springs}\\
\textit{United States} \\
\textit{and George Mason University, Fairfax, VA, United States}\\
clarelaney26@gmail.com} ~\\
\and
\IEEEauthorblockN{Andrew Barovic and Armin Moin\textsuperscript{*}}
\IEEEauthorblockA{\textit{Department of Computer Science} \\
\textit{University of Colorado Colorado Springs}\\
\textit{United States} \\
{\{abarovi2, amoin\}}@uccs.edu \\
\textsuperscript{*}Corresponding author
}
}
\maketitle

\begin{abstract}
Many users and contributors of large open-source projects report software defects or enhancement requests (known as bug reports) to the issue-tracking systems. However, they sometimes report issues that have already been reported. First, they may not have time to do sufficient research on existing bug reports. Second, they may not possess the right expertise in that specific area to realize that an existing bug report is essentially elaborating on the same matter, perhaps with a different wording. In this paper, we propose a novel approach based on machine learning methods that can automatically detect duplicate bug reports in an open bug repository based on the textual data in the reports. We present six alternative methods: Topic modeling, Gaussian Na\"ive Bayes, deep learning, time-based organization, clustering, and summarization using a generative pre-trained transformer large language model. Additionally, we introduce a novel threshold-based approach for duplicate identification, in contrast to the conventional top-k selection method that has been widely used in the literature. Our approach demonstrates promising results across all the proposed methods, achieving accuracy rates ranging from the high 70\%'s to the low 90\%'s. We evaluated our methods on a public dataset of issues belonging to an Eclipse open-source project.
\end{abstract}


\section{Introduction} \label{introduction}
Large open-source projects offer issue-tracking systems with open bug repositories, where developers and users can report the software defects they find or any new feature requests they may have. These reports are called \textit{bug reports}. However, some bug reports in the open bug repository are duplicates. They essentially describe the same software defect or enhancement request with some different phrasing. Nevertheless, it may not be evident to non-experts or even experts who may not be specialized in that specific area of the project that two or more bug reports are redundant (i.e., they are describing the same issue).

To make bug report processing and resolution more efficient, large open-source projects typically adopt a process called \textit{bug triage}. Similar to a hospital's emergency department or a combat zone, where due to limited resources, a triage nurse should first see patients or casualties to determine whether a healthcare provider should see them and how urgently this must occur, a small group of developers, called \textit{bug triagers} should examine every newly reported issue. They first determine whether the issue is a valid one. In some cases, issues are invalid since, for example, the user did not know how to use the software properly as a result of not reading the documentation. Invalid bug reports must be marked as INVALID. Furthermore, if a new issue is a duplicate of an existing bug report in the open bug repository, the issue must be labeled as DUPLICATE and must be linked to the existing instance. Otherwise, an issue that is valid and original should be assigned to a developer who has the availability to work on the issue and possesses the required expertise and interest in that area. These are all jobs of bug triagers, who should have a very good understanding of the open-source project.

Over the past two decades, (semi-)automated approaches based on various Artificial Intelligence (AI) methods, mainly Natural Language Processing (NLP) using Machine Learning (ML) or Information Retrieval (IR), have been proposed to assist in the bug triage process. For instance, Anvik et al. \cite{Anvik+2006} used ML methods to help in the bug assignment part of bug triage, whereas Sun et al. \cite{Sun+2011} deployed IR to assist in the duplicate bug report detection task. The overall goal of all the related work in this field has been to reduce the total maintenance and evolution cost of large open-source projects over time through Mining Software Repositories (MSR). 

In this work, we concentrate on automated duplicate bug report detection using MSR with a focus on software bug repositories. Note that redundant bug reports could unnecessarily increase the project's costs in two ways. First, they take the bug triagers' time to be detected. Second, if they remain undetected, which is not unlikely, they will lead to a waste of developers' time, as they may need to work on the same issue twice. Hence, increasing the effectiveness and efficiency of the duplicate bug report detection task is a priority in almost all large open-source software projects.

This paper proposes a novel approach to the automated detection of duplicate bug reports in large open bug repositories based on recent advances in NLP, including the Bidirectional Encoder Representations from Transformers (BERT) \cite{Devlin+2019} and the generative pre-trained transformers (GPT) \cite{OpenAI2023-GPT4} large language models (LLMs). To this aim, we analyze the natural language textual information in the summary/title, description, and comments of existing bug reports for a specific project. First, we create several topics and assign each bug report to one of them. This step is based on the work of Xia et al. \cite{Xia+2017} and is called topic modeling. Second, we use several similarity measures to find highly similar or potentially identical bug reports among those that belong to the same topic. Here, we propose five alternative methods, namely Gaussian Na\"ive Bayes, deep learning, time-based organization, clustering, and summarization.

This paper's contribution is twofold: First, it proposes a novel approach to automated duplicate bug report detection in which bugs are classified in a binary manner as duplicate or not duplicate. Second, in the case of duplicate bug reports, it helps find other similar bugs in the open bug repository. It provides an open-source prototype that enables other open-source projects using similar issue-tracking systems to deploy the proposed approach, thus benefiting from effective and efficient automated detection of duplicate bug reports.

This paper is structured as follows: Section \ref{background} provides some background information about open bug repositories and NLP. Further, we review the literature in Section \ref{related-work}. In Section \ref{proposed-approach}, we propose our novel approach and report on our experimental results in Section \ref{experimental-results}. Moreover, Section \ref{threats-to-validity} points out potential threats to validity. Finally, we conclude and suggest future work in Section \ref{conclusion-future-work}.

\section{Background} \label{background}
\subsection{Open bug repositories}
Bug repositories store bug reports for a software project. In open-source projects, there is typically an open bug repository where anyone who creates an account can file a bug report. This leads to many reports, particularly for large open-source projects with a significant user and developer community. For example, according to Anvik et al. \cite{anvik_coping_2005}, in 2005, one of the Mozilla projects' bug triagers had stated that almost 300 bugs would appear every day that needed triaging, which would be far beyond their project's resources to handle \cite{anvik_coping_2005}.

\begin{figure*}
    \centering
    \includegraphics[width=1\linewidth]{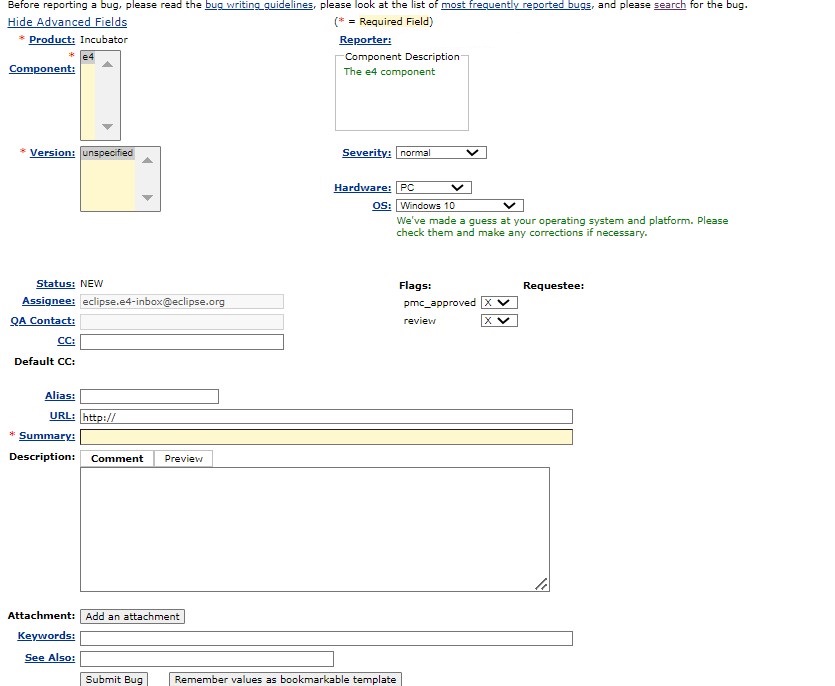}
    \caption{Reporting a new issue to the Eclipse Bugzilla repository}
    \label{fig:myBugzillaReport}
\end{figure*}

Bug reports include various free-form textual data fields (e.g., summary/title and description) and textually predefined data (e.g., product and component). The summary (or title) is usually a one-line overview of the issue. In contrast, the description is a detailed outline of the issue. Both summary and description are free-form since the bug reporter can write text in an unstructured manner for these fields. However, in the case of predefined fields, we typically have drop-down menus where the bug reporter can choose an option among the predefined set of choices. Figure \ref{fig:myBugzillaReport} depicts the user's view when reporting a new issue to the Bugzilla issue tracking system of the Eclipse open bug repository.

Note that it is crucial to preserve duplicate bug reports and link them to the existing (redundant) instances instead of simply dropping them or just marking them as duplicates but then leaving them as isolated instances. The reason that both labeling and linking them to each other is crucial is that hearing about the same issue from different perspectives, perhaps based on running or using the software on various platforms and with different levels of granularity and technical details, will typically help developers better comprehend, reproduce, analyze, and resolve the issue.

Last but not least, bug reports that triagers should exclude from the bug fixing process (i.e., sort of \textit{desk-reject}), including invalid and duplicate bug reports, occur quite frequently in large open-source ecosystems. For instance, over 2018-2021, up to 39\% of the Eclipse bug reports have been either duplicates or invalid reports \cite{jiang_does_2023}. Therefore, it is important to come up with automated approaches that can assist in detecting such cases. In this paper, we focus on duplicate cases.

\subsection{Natural Language Processing (NLP)}
NLP is a subfield of AI that is concerned with computers interpreting and reasoning on human-spoken languages (i.e., natural languages). Methods and techniques from other AI sub-disciplines, such as ML, are frequently used in NLP and closely related fields, namely natural language understanding and natural language generation.

An NLP pipeline often starts with data pre-processing. Pre-processing is a phase in which the data is put into a form that is easier for the computer to process. This could include tokenizing, stemming, and removing stop words. \textit{Tokenizing} is splitting a sentence into tokens by delimiters. Delimiters include spaces, commas, or punctuation. For instance, if we consider whitespace to be a delimiter, then tokens would be words. Further, \textit{stemming} is about transforming words that have multiple forms into a single form. For example, \textit{fly}, \textit{flew}, and \textit{flying} should all be converted into \textit{fly}. Moreover, stop words include some quite common words in text corpora, such as \textit{an}, \textit{a}, and \textit{the}. 

Additionally, various methods for text vectorization exist. Count vectorizer, TF-IDF, and Word2Vec are some well-established options. 

\section{Related Work} \label{related-work}
Computer-assisted bug triage in open-source software systems has been studied over the past decades. Various approaches and techniques, such as using machine learning for bug localization and automated bug assignment based on that \cite{MoinKhansari2010}, or automated bug assignment using information retrieval based on developers' profiles \cite{MoinNeumann2012}, were proposed.

Xia et al. \cite{Xia+2017} proposed a topic modeling approach to categorize bug reports. Their method, called Multi-feature Topic Modeling (MTM), was an extension of the popular Latent Dirichlet Allocation (LDA) method \cite{blei+2003}. They considered the textually predefined data fields of the bug reports (such as the product and component) in addition to the summary and description to improve the performance of their topic categorization method.

Sun et al.'s work \cite{Sun+2011} was a relatively early paper on duplicate bug report detection. They proposed an Information Retrieval (IR) function (a so-called \textit{REP function}) and used a combination of predefined and freeform textual data. The weights of their REP function were optimized during training using stochastic gradient descent. The larger weights would indicate that the respective feature could better distinguish between similar and non-similar bug reports. They also introduced an improved statistical IR technique, called \textit{BM25F ext}, to analyze the similarity of long textual bug report queries.

Nguyen et al. \cite{nguyen_duplicate_2012} used topic modeling and textual similarity (BM25F) to detect duplicate bugs. This approach improved the REP function's performance by up to 20\% in some cases. Their techniques were especially effective for bug reports that described the same issue with different technical terms.

Zhang et al. \cite{Zhang+2023} carried out a literature review of the previous work on automated duplicate bug report detection approaches. Experimental results from this review showed that the IR (REP-based) methods outperformed more complex Neural Network-based approaches. They also found that duplicate bug detection techniques that work well on older bug reports may not necessarily perform effectively on more recent bug reports. Moreover, they pointed out that duplicate bug detection techniques for one issue-tracking system may not work well on other issue-tracking systems.

Additionally, Zhang et al. \cite{zhang_cupid_2023} proposed a novel approach to duplicate bug reports by leveraging GPT, which summarized long reports before they were classified using the REP function created by Sun et al. \cite{Sun+2011}. This improved the \textit{Recall Rate@10} of the REP by up to 8.7\%. We also deploy GPT for summarization in one of our methods.  

Moreover, Xie et al. \cite{Xie+2018} and He et al. \cite{He+2020} trained Convolutional Neural Network (CNN) models to identify duplicate bugs. Finally, in contrast to Zhang et al.'s conclusion \cite{Zhang+2023}, Jiang et al.'s study \cite{jiang_does_2023} found that deep learning techniques were better at classifying bugs as duplicates or non-duplicates (i.e., binary) than traditional NLP approaches based on IR. However, deep learning approaches performed worse at creating a list of possible duplicates of a given bug report. 

Therefore, based on the results of the latter study \cite{jiang_does_2023}, we deploy deep learning to predict whether a specific bug report is likely to be a duplicate of an existing bug report. However, we use traditional IR-based methods to obtain a list of potentially duplicates of a particular issue identified as duplicate.

\section{Proposed Approach} \label{proposed-approach}
\subsection{Topic Modeling}
We first conduct topic modeling using LDA \cite{blei+2003}. Given a set of $n$ bug reports, $\{r_1, ..., r_n\}$, in the open bug repository of an open-source project, we deploy LDA to automatically create a number of topics $\{t_1, ..., t_k\}$ where $k \leq n$ and $\forall x \in \{r_1, ..., r_n\} $ $\exists $ t $\in \{t_1, ..., t_k\} $ such that $topic(x) == t$. This step can make our following tasks more efficient.

\subsection{Duplicate bug report detection}
Given a set of $n$ bug reports, $\{r_1, ..., r_n\}$ and their corresponding boolean flags $is\_dup(r_x)$ $\forall x \in \{1, ..., n\}$ indicating whether they are duplicates or not, we want to be able to predict the correct duplicate flag for a new bug report, say $is\_dup(r_{n+1})$ automatically, based on the textual information that is contained in this new bug report. To this aim, we deploy an ML model per bug report topic $t \in \{t_1, ..., t_k\}$ and train it with the textual information and the boolean duplicate flags assigned to the bug reports $\{r_j, ... r_p\} \subset \{r_1, ..., r_n\}$, where $\forall x \in \{r_j, ... r_p\}$ $ status(x) == RESOLVED\_FIXED $ $\land $ $topic(x) == t$.

The reason that we train our ML model only on bug reports that are labeled as \textit{RESOLVED\_FIXED} is that we want to make sure our classifier will not see much noise and incorrect labels as opposed to \textit{reasonable} labels. For instance, a bug report's duplicate flag may not have been assigned accurately at the beginning of its lifecycle. However, the assumption is that by its resolution time, the flag has been corrected if necessary.

Given a new bug report $r_{n+1}$, for which $status(r_{n+1}) == NEW$, we first find the most relevant topic $t \in \{t_1, ..., t_k\}$. The trained ML model associated with that topic, say $ML_t$ will be capable of determining the duplicate status of $r_{n+1}$, which is $dup(r_{n+1}) \in \{True, False\}$.

\subsection{List of duplicates generation}
Given an open bug repository with $n$ bug reports, $\{r_1, ..., r_n\}$, we want our IR-based method to use a similarity measure, namely the \textit{Cosine} similarity measure, to automatically generate a list of highly similar (thus potentially duplicate/redundant) bug reports $\{r_h, ... r_m\} \subset \{r_1, ..., r_n\}$ to a new bug report $r_{n+1}$, where $dup(r_{n+1}) == True $ $ \land $ $\forall x \in \{r_h, ... r_m\}$ $ [ dup(x) == True $ $ \land $ $ \exists $ $ \delta $ such that $ Cosine\_sim(x, r_{n+1}) > \delta] $.

In Section \ref{experimental-results}, we explain how we convert textual information in bug reports to efficient numeric representations, known as \textit{vector embeddings}. Assuming two bug reports $r_a$ and $r_b$ are represented with vectors $\mathbf{A}$ and $\mathbf{B}$, respectively, the Cosine similarity measure for these two vectors, which indicates the similarity of bug reports $r_a$ and $r_b$, is defined as follows:
\[
\text{Cosine\_sim}(\mathbf{A}, \mathbf{B}) = \frac{\mathbf{A} \cdot \mathbf{B}}{\|\mathbf{A}\| \|\mathbf{B}\|}
\]
where \(\mathbf{A} \cdot \mathbf{B}\) is the dot product of the two vectors, and \(\|\mathbf{A}\|\) and \(\|\mathbf{B}\|\) are their magnitudes (i.e., Euclidean norms). More similar word vector embeddings have a smaller angle between them in the vector embeddings space. Thus, the cosine of the angle between them is larger (i.e., closer to 1). A Cosine similarity of 1 means 100\% similarity.

\section{Experimental Results} \label{experimental-results}

\subsection{Research method}
We deploy the empirical research method called Repository Mining \cite{acm-empirical-methods-repo-mining}. In the following, we elaborate on the dataset used for evaluation and the evaluation metrics. We then illustrate our experimental results.

\subsection{Dataset}
\begin{figure}
    \centering
    \includegraphics[width=0.6\linewidth]{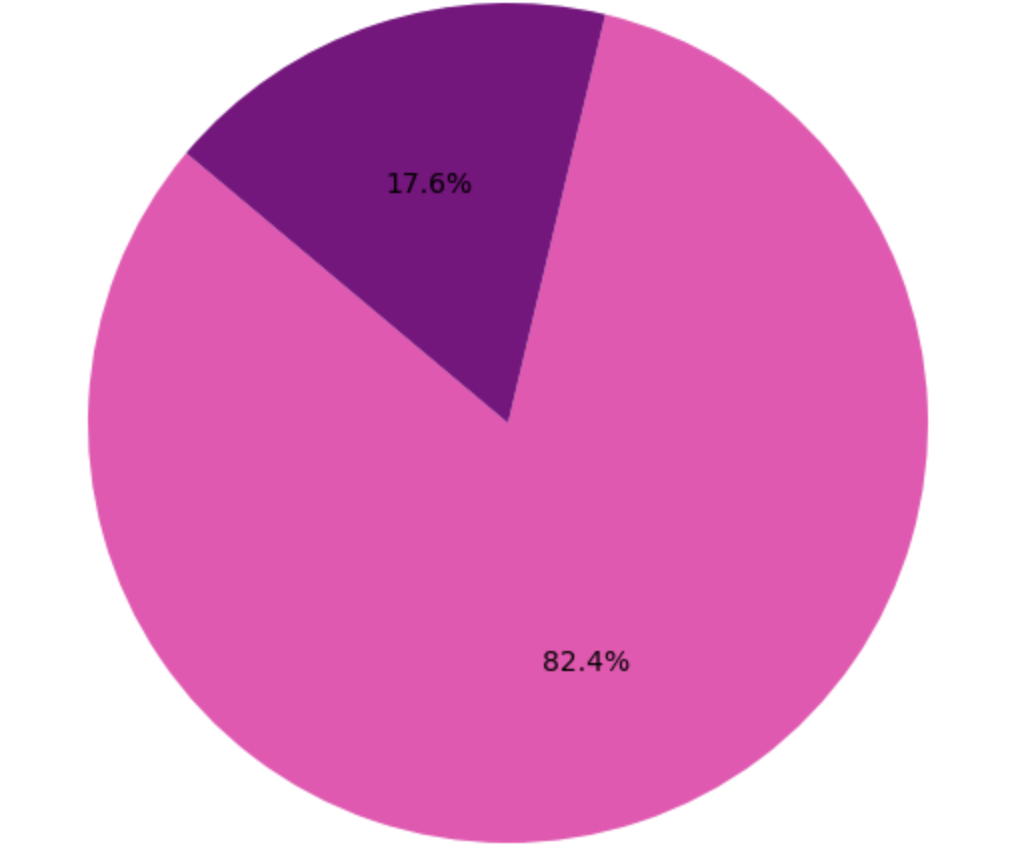}
    \caption{Duplicate (17.6\%) vs. non-duplicate (82.4\%) bug reports in our dataset}
    \label{fig:dupGraph}
\end{figure}

We used a dataset on GitHub called BugHub \cite{datasethomepage}. This is an open reference dataset that contains a collection of free-text bug reports for research on duplicate issue identification. The bug reports are from 2002 to 2013 and belong to the Eclipse Bugzilla repository. We selected around 75,000 bug reports from this dataset. As shown in Figure \ref{fig:dupGraph}, duplicate bug reports constitute 17.6\% of our experimental data. 

\subsection{Evaluation metrics}
Recall that we carry out two tasks. First, we predict whether a bug report should be labeled as a duplicate bug report or not. This is a binary classification (supervised ML) task. Second, we generate a list of potentially duplicate bug reports for a specific bug report that is labeled as a duplicate (called the issue of interest) by finding the most similar bug reports to the issue of interest. This is an IR task that is concerned with text similarity detection. Consequently, we have different ways of evaluating the performance of our proposed approach for the two mentioned tasks. In both of them, we report the Accuracy, Precision, Recall, and F1-measure (also known as F1 Score), which are typical evaluation metrics in the literature. Assuming we have n classes, these metrics are defined in Equations \ref{eq:accuracy}, \ref{eq:precision}, \ref{eq:recall}, and \ref{eq:f1} below, where TP, TN, FP, and FN indicate the number of true positive, true negative, false positive, and false negative instances, respectively.

\begin{equation}
\begin{split}
Accuracy = \\
\frac{\sum_{i=1}^{n} TP_i + \sum_{i=1}^{n} TN_i}{\sum_{i=1}^{n} TP_i + \sum_{i=1}^{n} FP_i + \sum_{i=1}^{n} TN_i + \sum_{i=1}^{n} FN_i}
\end{split}
\label{eq:accuracy}
\end{equation}

\begin{equation}
Precision = \frac{1}{n} \sum_{i=1}^{n} \frac{TP_i}{TP_i + FP_i}
\label{eq:precision}
\end{equation}

\begin{equation}
Recall = \frac{1}{n} \sum_{i=1}^{n} \frac{TP_i}{TP_i + FN_i}
\label{eq:recall}
\end{equation}

\begin{equation}
F1-measure = \frac{1}{n} \sum_{i=1}^{n} \frac{2 \cdot \frac{TP_i}{TP_i + FP_i} \cdot \frac{TP_i}{TP_i + FN_i}}{\frac{TP_i}{TP_i + FP_i} + \frac{TP_i}{TP_i + FN_i}}
\label{eq:f1}
\end{equation}

Concerning the IR task, we conduct two different kinds of evaluation. First, we follow the literature and adopt the typical evaluation technique, called \textit{Top-K-based} evaluation, that is widely used in prior work. We do this to be able to compare our experimental results with them. This evaluation method ranks bug reports and considers a fixed number of top-ranked ones, say top-k, regardless of their similarity level. Our experimental results for the IR task using the top-k method can be found in table \ref{tab:topKresults}. Note that prior works in the literature only reported the recall but not the precision, and we know that there is often a trade-off between the two metrics.

Second, we use a different evaluation technique, which we find reasonable, and name it \textit{Threshold-based} evaluation. In this method, we consider bug reports above a certain similarity threshold to be duplicates of each other. We consider the results to be a success if at least one bug with a similarity above the threshold was indeed a duplicate in the ground truth or if the method correctly identified a non-duplicate and marked it as such. 

\begin{figure*}
    \centering
    \includegraphics[width=1\linewidth]{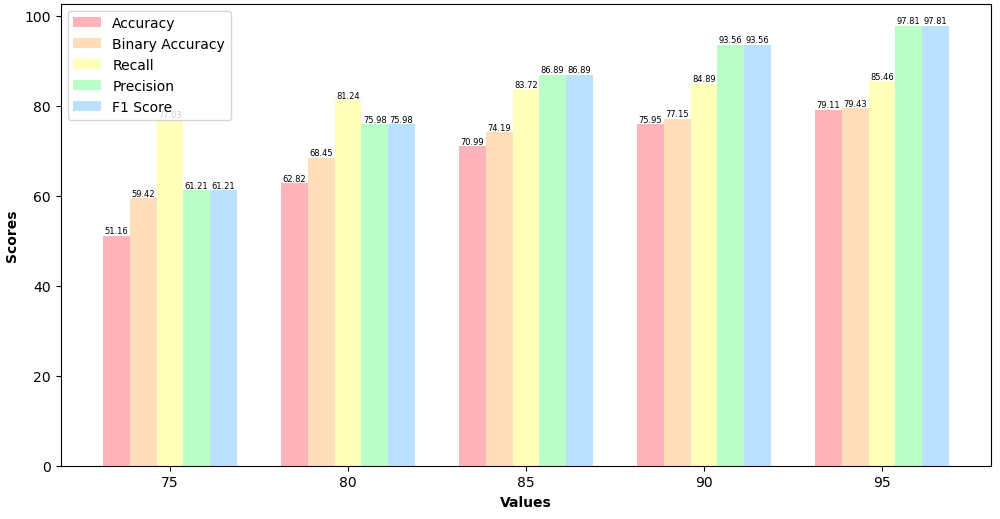}
    \caption{Testing different similarity thresholds}
    \label{fig:thresholds}
\end{figure*}

Our experimental results for the IR task using the threshold-based evaluation technique are available in \ref{tab:thers_exp_results}. As illustrated in Figure \ref{fig:thresholds}, a more restrictive (i.e., higher) similarity threshold value performs better across all metrics. Unless otherwise specified, our results are averaged across all seven topics.

Below, we briefly describe how the Binary Accuracy, Accuracy, Precision, and Recall are defined for this task's evaluation.

\paragraph{Binary Accuracy}
This measures the accuracy of the approach in determining whether a given bug report is a duplicate or not. This evaluates the performance of the binary classification task.

\paragraph{Accuracy}
This measures how accurately the approach identifies the correct duplicate bug IDs or returns none if there are no duplicates. This evaluates the effectiveness of generating a list of duplicate bug reports for a particular issue (i.e., an issue of interest).

\paragraph{Recall-Rate@K}
The Recall-Rate@K is defined as the proportion of the number of relevant items in the top-k to the total number of relevant items. Relevant in this context means duplicate bug reports. This is the metric that has been adopted in the literature by the prior work.

\paragraph{Precision-Rate@K}
The Precision-Rate@K is defined as the proportion of the number of duplicates in the top-k to the total number of items in the top-k.

\begin{table*}[h]
    \centering
    \caption{Experimental results for the IR task using the top-k evaluation method compared to the state of the art}
    \label{tab:topKresults}
    \begin{tabular}{|l|c|c|c|}
        \hline
        & \textbf{Our Approach} & \textbf{REP} \cite{Sun+2011} & \textbf{DBTM (State of the Art)} \cite{nguyen_duplicate_2012}\\
        \hline
        RecallRate@5 & 15.13\% & 64\% \cite{Sun+2011} & 76\% \cite{nguyen_duplicate_2012} \\
        \hline
        RecallRate@10 & 17.95\% & 70\% \cite{Sun+2011} & 82\% \cite{nguyen_duplicate_2012}\\
        \hline
        RecallRate@15 & 19.31\% & 73\% \cite{Sun+2011} & 85\% \cite{nguyen_duplicate_2012}\\
        \hline
        RecallRate@20 & 20.38\% & 77\% \cite{Sun+2011} & 87\% \cite{nguyen_duplicate_2012}\\
        \hline
        PrecisionRate@5 & 79.51\% & Not available & Not available \\
        \hline
        PrecisionRate@10 & 79.52\% & Not available & Not available \\
        \hline
        PrecisionRate@15 & 79.53\% & Not available & Not available \\
        \hline
        PrecisionRate@20 & 79.55\% & Not available & Not available \\
        \hline
        Mean Avg Precision & 79.53\% & 58.96\% \cite{Sun+2011} & Not available \\
        \hline
    \end{tabular}
\end{table*}

\begin{table*}[h!]
    \centering
    \caption{Experimental results for the IR task using the threshold-based evaluation method (rounded)}
    \label{tab:thers_exp_results} 
    \begin{tabular}{|l|c|c|c|c|c|c|c|}
        \hline
        & No Topic Modeling & 85\% threshold & 90\% threshold & 95\% threshold & Time sorted by quarter & Cluster & GPT summary \\
        \hline
        Accuracy & 77 & 70 & 76 & 79 & 83 & 81 & 92 \\
        \hline
        Binary Accuracy & 79 & 74 & 77 & 79 & 83 & 81 & 92 \\
        \hline
        Recall & 87 & 83 & 85 & 85 & 85 & 85 & 94 \\
        \hline
        Precision & 94 & 87 & 94 & 98 & 100 & 100 & 99 \\
        \hline
        F1 Score & 90 & 85 & 89 & 98 & 92 & 92 & 96 \\
        \hline
    \end{tabular}
\end{table*}

\subsection{Training and testing data segregation}
The BugHub Eclipse dataset \cite{datasethomepage} contains 68,124 bug reports. The specific dataset breakdown in terms of duplicates vs. non-duplicates can be viewed in Figure \ref{fig:dupGraph}. The BERT, Multi-Layer Perceptron (MLP), and Na\"ive Bayes models we trained each use this dataset in a different way. The MLP model uses a 75\% training and 25\% test set and performs all calculations on the CPU. The BERT model has a different dataset split, as we also deployed a validation set. To create this validation set, the testing set is first given 25\% of the dataset. This initial assignment bias is what creates the discrepancy in the non-duplicate and duplicate split of the data shown in Tables \ref{tab:MLP}, \ref{tab:BERt}, and \ref{tab:NaiveBayes}. Once the testing set is assigned 25\% of the dataset, the remaining bug reports are split between validation and training; 80\% of the remaining dataset was used for training, with 20\% of the remaining dataset being used for validation.

\subsection{The baseline approach: Cosine similarity}
For our baseline, we use the Cosine similarity measure. Bug reports above 85\% similar are considered duplicates. We compare all bug reports with all other bug reports within the dataset.

\subsubsection*{Word embeddings}
We use the sentence transformer library in Python to embed the bug reports. An embedding is a way to represent a word numerically in NLP. The embeddings of the bug reports' title/summary and description are from the sentence\_transformers library. We use the paraphrase-MiniLM-L6-v2 model. Based on these embeddings, we calculate the Cosine similarity.  

\subsection{Topic modeling and Cosine similarity}
To improve the baseline, we implement the LDA topic modeling as well as the Cosine similarity methods to categorize bug reports as duplicate or non-duplicate and output the similar duplicate bug report IDs. 

\subsubsection*{Latent Dirichlet Allocation (LDA)}
\paragraph*{Preprocessing}
We merge each bug's title/summary and description into a single column named full\_text\_data. We then tokenize this column, breaking it down into individual words. After tokenization, we remove irrelevant tokens, such as stop words, words that appear in over 90\% of the bug reports, and words that occur less than twice.

\paragraph*{Count vector}
Next, we create a vector indicating how often each token occurs in each bug report. The length of this vector is equal to the vocabulary size. 

\paragraph*{Document-term matrix}
After that, we create a sparse matrix in which each row represents a bug report, and each column represents a word in the vocabulary. We fill in our matrix with the previously calculated count vector. 

\paragraph*{Term co-occurrence matrix}
LDA creates a term or word co-occurrence matrix, which shows how often words appear together. This model assumes that words that co-occur should be in the same topic. The algorithm then uses both the document-term matrix and the term co-occurrence Matrix to generate topic distributions. 

\paragraph*{Topic distributions}
A topic distribution is a list of the most likely words associated with each topic and their probabilities. It is chosen randomly and improved iteratively. 

\paragraph*{Gibbs sampling} 
Gibbs sampling assigns words to each topic through an iterative process:
\begin{equation}
P(z_{i} = k \mid \mathbf{z}_{-i}, \mathbf{w}) \propto (n_{d,k}^{-i} + \alpha_k) \frac{n_{k,w_i}^{-i} + \beta}{n_{k}^{-i} + V\beta}
\end{equation}
The first fraction represents how important a topic is in a document, and the second fraction represents how often a word is assigned to a given topic throughout all bug reports. 

\begin{figure*}
    \centering
    \includegraphics[width=0.85\linewidth]{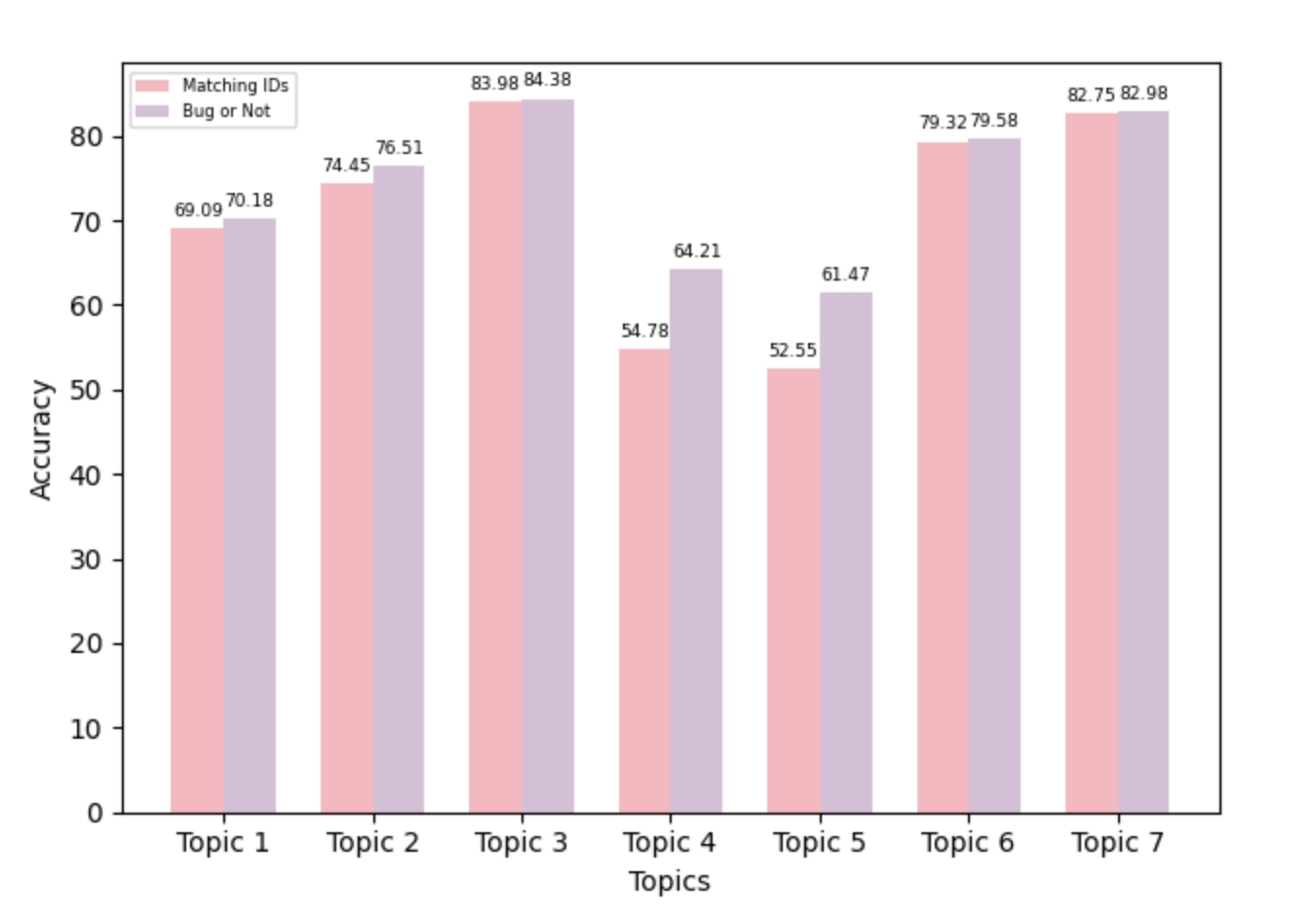}
    \caption{85\% similarity results for each topic}
    \label{fig:85sim}
\end{figure*}

\paragraph*{The number of topics and the similarity threshold} 
The LDA method outputs $k$ topics and assigns each bug report to a topic. We categorize the bug reports into seven topics. This number can be changed. For the current dataset, $k=7$ works well. Then, we compare the Cosine similarity of bug reports within the same topic. For our initial experimentation, we consider an 85\% similarity threshold. Our results for 85\% similarity can be found in Figure \ref{fig:85sim}. In Figure \ref{fig:thresholds}, we report the average performance across the topics.

\subsubsection*{Deep learning}
We also experiment with deep learning techniques to categorize bug reports as duplicates or non-duplicates. We report on the results using the following models: MLP (see Table \ref{tab:MLP}) and BERT (see Table \ref{tab:BERt}).

MLP is a fully connected feed-forward neural network. In this case, our features are the title/summary and description embeddings using the TF-IDF Vectorizer, which uses the inverse document frequency and the term frequency. We deploy a ReLU activation function. Surprisingly, an MLP model that just had the component as a feature performed better than one that had the textual data as a feature. An MLP model with both the component and the free-form textual data performed similarly to one with just the free-form textual data. 

BERT (Bidirectional Encoder Representations from Transformer) is a transformer model that we use to classify our bug reports as duplicate or non-duplicate. We begin by using BERT to tokenize the title/summary and description of the bug reports. We use ten epochs to train the model on our dataset. We utilize the AdamW optimizer from hugging face for our model.

\subsection{Gaussian Na\"ive Bayes}
Gaussian Na\"ive Bayes is a classical machine learning method. It assumes that the features are independent of each other and that they have a normal distribution. Note that these assumptions do not hold for our data. A Gaussian Na\"ive Bayes model was also tested, which performed only slightly worse than the MLP model in every metric. This model also deploys a word2vec vectorizer.

\subsection {Time seperated}
We experiment with splitting the data into different time sections and comparing the Cosine similarity of bug reports submitted within similar time frames. For our initial experimentation, we split bug reports by quarters (i.e., 3-month) periods. The hypothesis is that very similar or essentially the same issues would be submitted to the bug repository around the same time. We replaced topic modeling with time separation. We compared similar bug reports within the same 3-month period and categorized those with an 85\% or higher Cosine similarity as duplicates. We created 27 different 3-month time frames. This method performed better than topic modeling, confirming our hypothesis that similar or the same issues happen to be reported around the same time period. Table \ref{tab:thers_exp_results} shows the experimental results.

\subsection {Clustering}
We experimented with K-means clustering as an alternative to topic modeling. We set the hyperparameter of K to 10. Table \ref{tab:thers_exp_results} shows the experimental results.

\subsubsection{K-means}
We used K-means clustering as an alternative to topic modeling. We created ten different clusters and compared the bugs within each cluster. K-means clustering is an algorithm that starts by randomly initializing. K centroids and iteratively updates them to minimize the Euclidean distance between the data points and the centroids. We chose a hyperparameter of 10 using the elbow method. We tested the even K values from zero to twenty and then computed the sum of squared distances between each point and its nearest centroid, known as its inertia. As the number of clusters increases, the inertia decreases because the distance between a point and its nearest centroids becomes smaller. We created a graph with the y-axis representing the inertia and the x-axis representing the number of clustering. Then, by visual inspection, we identify the bend or \lq{}elbow\rq{} in the plot. This is the point at which the rate of decrease in the inertia slows significantly as we increase the number of clusters. This method should pick a point that balances the model's complexity and performance. Table \ref{tab:thers_exp_results} shows the experimental results.

\subsection{GPT Summarization}
We experiment with using the GPT 3.5 turbo API to summarize the bug reports before using the Cosine similarity to compare them. We could only summarize about 2,000 bug reports due to the computational expenses of this method. By summarizing the bug reports, we hoped that the model would phrase the reports more simply so that they would be easier to compare. We also wanted summaries to remove hyperlinks that would be hard to embed accurately and potentially skew our Cosine similarity values. We prompted the model with the below paragraph: 
\begin{quote}
"I want you to act as a collaborator for maintaining bug reports from a large open source software project. Your job is to rephrase the reports to avoid repetition while keeping semantic meaning. The goal of this process is to help filter duplicates. Please only output the summary of the bug report."
\end{quote}
We then used the Cosine similarity with a threshold of 95\% to compare and determine the duplicate bug ID based on the summarized descriptions. This heavily improves the accuracy, binary accuracy, and recall of the topic modeling and Cosine similarity approach (using the same threshold value). The precision and the F1 score stayed the same. Further experimentation is needed to validate this result, as the summarized approach was implemented with fewer bugs than the topic modeling approach. This idea was inspired by Zhang et al.'s work \cite{zhang_cupid_2023}, which also used summarization and a duplicate bug classification function. Table \ref{tab:thers_exp_results} shows the experimental results.

\begin{table}[h!]
    \centering
    \caption{Experimental results using the MLP model}
    \begin{tabular}{|c|c|c|c|c|}
        \hline
        Class & Precision & Recall & F1-Score & Support \\
        \hline
        0 & 0.84 & 0.87 & 0.85 & 14,032 \\
        1 & 0.26 & 0.21 & 0.23 & 2,999 \\
        \hline
        \multicolumn{5}{|c|}{Average Values} \\
        \hline
        Micro & 0.55 & 0.54 & 0.54 & 17,031 \\
        Weighted & 0.74 & 0.75 & 0.74 & 17,031\\
        \hline
        \multicolumn{5}{|c|}{Accuracy: 0.75} \\
        \hline
    \end{tabular}
    
    \label{tab:MLP}
\end{table}

\begin{table}[h!]
    \centering
    \caption{Experimental results using the BERT model}
    \begin{tabular}{|c|c|c|c|c|}
        \hline
        Class & Precision & Recall & F1-Score & Support \\
        \hline
        0 & 0.85 & 0.74 & 0.79 & 14,031 \\
        1 & 0.23 & 0.36 & 0.28 & 3,000 \\
        \hline
        \multicolumn{5}{|c|}{Average Values} \\
        \hline
        Micro & 0.54 & 0.55 & 0.54 & 17,031 \\
        Weighted & 0.74 & 0.68 & 0.70 & 17,031\\
        \hline
        \multicolumn{5}{|c|}{Accuracy: 0.68} \\
        \hline
    \end{tabular}
    
    \label{tab:BERt}
\end{table}

\begin{table}[h!]
    \centering
    \caption{Experimental results using the Na\"ive Bayes model}
    \begin{tabular}{|c|c|c|c|c|}
        \hline
        Class & Precision & Recall & F1-Score & Support \\
        \hline
        0 & 0.83 & 0.85 & 0.84 & 14,032 \\
        1 & 0.22 & 0.20 & 0.21 & 2,999 \\
        \hline
        \multicolumn{5}{|c|}{Average Values} \\
        \hline
        Micro & 0.53 & 0.52 & 0.53 & 17,031 \\
        Weighted & 0.73 & 0.74 & 0.73 & 17,031\\
        \hline
        \multicolumn{5}{|c|}{Accuracy: 0.74} \\
        \hline
    \end{tabular}
    \label{tab:NaiveBayes}
\end{table}

\section{Threats to Validity} \label{threats-to-validity}
The exact results for the recallRate@K for the REP and DBTM function are unknown, so we had to look at the chart and estimate the values. We could not replicate the results of the REP and DBTM functions \cite{Sun+2011} on the dataset we used as the code is not public. The precision results for the REP function are not included in their work, which is why that section of Table \ref{tab:topKresults} has been left empty. The REP method had a Mean Average Precision, which our work compares to, but in either case, the precision values of the methods are not provided at specific K values. 

Furthermore, the results of this study are based on the chosen bug dataset. It may not be representative and generalizable to other software projects and bug repositories or datasets. Further investigation will b required to see how our methods will perform in other projects or for other datasets.

Finally, the reported results for GPT summarization in Table \ref{tab:topKresults} have been achieved on a subset of the dataset as mentioned. Therefore, it may not be comparable to the rest of the table.

\section{Conclusion and Future Work} \label{conclusion-future-work}
In this paper, we have proposed a novel approach to the automated detection and linking of duplicate bug reports in large open-source projects. We proposed several new methods and techniques based on natural language processing, information retrieval, and machine learning. The multi-layer perception neural network model outperformed other methods. Considering the bug report timelines has proven to be very useful. Also, bug report summarization using GPT seems to be a promising direction that should be studied further in future studies. We hope to experiment with changing the number of topics used in our topic modeling in the future. We also want to improve the topic modeling from the standard LDA used in these experiments to more sophisticated methods.

\section*{Software and Data Availability}
The prototype is available under a permissive open-source license at \url{https://github.com/qas-lab/LaneyREU}. The data used for the evaluation are also publicly available at \cite{datasethomepage}.

\section*{Acknowledgment}
This material is based upon work supported by the U.S. National Science Foundation (NSF) under Grant No. 2349452. Any opinions, findings, conclusions, or recommendations expressed in this material are those of the authors and do not necessarily reflect the views of the NSF.

\bibliography{refs}
\bibliographystyle{IEEEtran}

\end{document}